\documentclass[conference]{IEEEtran}
\IEEEoverridecommandlockouts
\usepackage{cite}
\usepackage{amsmath,amssymb,amsfonts}
\usepackage[noend]{algpseudocode}
\usepackage{svg}
\usepackage{graphicx}
\usepackage{textcomp}
\usepackage{xcolor}
\usepackage{listings}
\usepackage{caption}
\usepackage{pdfpages}
\usepackage{pythonhighlight}

\def\BibTeX{{\rm B\kern-.05em{\sc i\kern-.025em b}\kern-.08em
    T\kern-.1667em\lower.7ex\hbox{E}\kern-.125emX}}
\begin{document}

\title{Opening the Black Box: Performance Estimation during Code Generation for GPUs}

\author{\IEEEauthorblockN{Dominik Ernst\IEEEauthorrefmark{1} (dominik.ernst@fau.de), Georg Hager\IEEEauthorrefmark{1}, Markus Holzer\IEEEauthorrefmark{2}, Matthias Knorr and Gerhard Wellein\IEEEauthorrefmark{1}} \IEEEauthorblockA{\IEEEauthorrefmark{1} NHR@FAU, Friedrich-Alexander-Universität Erlangen-Nürnberg, Germany} \IEEEauthorblockA{\IEEEauthorrefmark{2}Chair for System Simulation, Friedrich-Alexander-Universität Erlangen-Nürnberg, Germany}}


\maketitle

\begin{abstract}
  Automatic code generation is frequently used to create implementations of algorithms
  specifically tuned to particular hardware and application parameters.
The code generation process involves the selection of
adequate code transformations, tuning parameters, and parallelization
strategies. To cover the huge search space, code generation
frameworks may apply time-intensive autotuning, exploit
scenario-specific performance models, or treat performance as an
intangible black box that must be described via machine learning.

This paper addresses the selection problem by identifying the relevant
performance-defining mechanisms through a performance model coupled
with an analytic hardware metric estimator. This enables a quick
exploration of large configuration spaces to identify highly efficient
candidates with high accuracy.

Our current approach targets memory-intensive GPGPU applications and
focuses on the correct modeling of data transfer volumes to all levels
of the memory hierarchy.  We show how our method can be coupled to the
``pystencils'' stencil code generator, which is used to generate kernels for a
range four 3D25pt stencil and a complex two phase fluid solver based on the
Lattice Boltzmann Method. For both, it delivers a ranking that can be used
to select the best performing candidate.

The method is not limited to stencil kernels, but can be integrated into any
code generator that can generate the required address expressions.
\end{abstract}


\section{Introduction}

\subsection{Spoilt for Choice in a Vast Configuration Space}

Code Generation is a programming technique that promises
increased developer productivity through more abstract program
representations and specialization to both the application scenario
and the hardware.

For the same algorithm specification, the code generation process
can make different decisions, leading to a large variety of
possible outcomes. These decisions could be different
parallelization variants, the fusion, ordering, and unrolling of loops, or
different blocking or tiling schemes, together with the associated
tuning parameters.  The influence of these decisions on the
performance is often not obvious, and therefore a common method to
find the best configuration is to use autotuning with a more or less
exhaustive search of the configuration space.

The evaluation of each configuration requires code generation and compilation
and benchmarking of the executable, which can take impractically long
for a large configuration space.

A way out of this dilemma is to stop treating generated code and its
performance as an intangible black box but to make accurate
predictions instead about the resulting performance. This is
accomplished by a metric estimator that computes the inputs for
a performance model.

The main advantage of this approach is a much quicker evaluation time
compared to the generation-compilation-benchmark cycle.  We design the
hardware metric estimator with two goals in mind: It must only require
high-level features from the code generator that are already available
before the actual source code text is generated, and it must be
quickly evaluable.

Since there is no actual code execution in this process, it does not
require the target hardware, saving valuable resources especially
when high-end GPUs are involved. It also
enables the performance comparison of different GPU models, including
hypothetical GPUs for architectural exploration.
A performance model can not just evaluate different code configurations
on different hardware, but it can also grant insight into why the
performance is the way it is.  Knowing the performance-limiting
factors can inform further development of both algorithms and code
generation capabilities.

\subsection{Pystencils}

We show the utility of our performance estimator and how it
can be used in a code generator using the open-source python library \textit{pystencils} \cite{phasefields} as an example. With \textit{pystencils}, mathematical models can be directly described in an abstract representation. It is possible to automatically derive essential numerical schemes like the finite difference or the finite volume method from the abstract representation. Thus, complex mathematical models like multi-phase solidification models \cite{phasefields} can be described in a high-level latex-like representation close to usual descriptions in the literature. Furthermore, other packages like the Lattice-Boltzmann-Method (LBM) code generation framework lbmpy \cite{lbmpy1,lbmpy2} build on top of \textit{pystencils} to derive highly optimized numerical schemes for solving flow problems. From the numerical schemes, \textit{pystencils} then generates low-level C/OpenMP code to target CPUs or CUDA/OpenCL for GPUs.

As an example, consider the following 2D4Pt stencil:
\begin{python}
dst[0,0] = (src[1, 0] + src[-1, 0] +
            src[0, 1] + src[0, -1]) / 4
\end{python}

This example consists of a single assignment with relative field access to the destination and the source field dst and src.
\textit{Pystencils} then lowers this representation to its target dependent intermediate representation, where the relative field accesses are replaced with expressions computing the referenced addresses based on the thread coordinates, and the iteration space is realized either with explicit loops or mapped over hardware thread coordinates.
For example, for the access \pyth{src[1,0]} to the left neighbor, the referenced address would be computed on a GPU by this expression:

\begin{python}
  src_W = src +
        (threadIdx.x + blockIdx.x*blockDim.x + 1) +
        (threadIdx.y + blockIdx.y*blockDim.y) * w
\end{python}

The address expressions must contain only the base address of the field, and the thread and block coordinates as free fields.
Our performance estimator takes these address expressions, together with the launch configuration, field sizes and field alignments as the only high level information  required from a code generator.
Our estimator is therefore not specific to \textit{pystencils}, but can be integrated with any code generation framework that is able to generate these artifacts.

The requirements on the address expressions also introduces limitations to the applicability of our performance estimator.
First, variables like grid sizes need to be known at generation time and indirect addressing is not supported.
It possible to work around this using representative proxy values.

Second, control flow needs to be fully resolved beforehand. For very likely or unlikely, branches, it is often possible to simplify the control flow and still get a meaningful result.
The frequent \pyth{if(tidx >= N ) return 0;} at the top of many GPU kernels, is false for almost all grid points and thus can be dropped without changing the result.

Another example are grid stride loops, where each thread would iterate over many points of the computational domain.
In this case, the code generator can emit the address expressions for just use one or a few iterations, and normalize the performance to the amount of work performed in these loop iterations.

\subsection{Contributions}

In this paper, we make the following significant contributions.  We
extend the roof{}line performance model for GPUs by two additional
limiters related to cache bandwidth.  We present a method to estimate
the cache hierarchy data volume metrics required for this
model based on the address expressions used to compute the memory
addresses referenced by a GPU program.  We propose the combination of
the model and the metric estimator for  use by code
generators, which can generate these address expressions and use the
combination to classify different code generation options and run
configurations.  In this context, we demonstrate the usefulness
of the method for distinguishing badly- from well-performing configurations and
identifying which type of configuration performs best on a V100 GPU on
two different, challenging applications implemented in the stencil
code generation framework \textit{pystencils}.

\subsection{Related Work}

There are numerous code generation frameworks for
stencils on GPUs.  These frameworks use either autotuning, or a performance model
that is specific to their requirements.
For example, the authors of \cite{libra} use a model to optimize the
computation/register ratio, which is important for the class of
stencils they are targeting.  In \cite{an5d},  a standard roof{}line model
with a fixed, theoretical memory volume is used for a full exploration of the configuration space, followed by benchmarking the top five candidates.

Similarly to our strategy, the framework LIFT \cite{lift1,lift2} extracts low-level features from an intermediate representation (IR),
and then uses a machine learning approach to predict performance based on the
extracted code features.

The roof{}line model \cite{roofline_classic}, owes its popularity and wide
applicability to its simplicity, as it uses just two
performance limiters that apply to any architecture: 
memory bandwidth and peak floating-point performance.
These have been extended with cache-related limiters in the
\textit{hierarchical roof{}line model} \cite{roofline_hierarchical}
where data transfers volumes measured via hardware performance counters are
used to determine how close each memory level's bandwidth is to being a limiter.
NVIDIA's own \textit{nvvp} and \textit{nsight} profiling
applications do not just measure hardware quantities using performance
counters but also evaluate the measured results by comparing them
with the maximum possible rates.  This could be regarded as a multi-limiter
performance model that includes every limiter that is
measurable with performance counters.

In order to make a prediction, performance models focused on the memory hierarchy require application specific inputs in the form of the data volumes transferred in
the memory hierarchy.

One work that attempts to estimate these metrics for GPUs is
\cite{reuse}. They put GPU simulator traces into a reuse-distance-based
model with time stepping that can also resolve timing-based effects, which
is potentially more precise in determining memory hierarchy data volumes than
our approach. They do however employ much more involved simulations and require
GPU execution traces.

The work presented in \cite{analyticalstencilmodel} targets the same usage
scenario and uses a similar performance model to ours. It also describes
capacity miss effects probabilistically and performs a similar evaluation
to ours.  The biggest difference is in using analytical formulas specific to a particular
stencil to compute the memory volumes, which limits the applicability
to stencils, whereas our estimator does not have such a limitation.

\section{Theory}

\subsection{Performance Model}

The naive roof{}line model is often imprecise because
frequently, applications can saturate neither
of the two performance limits. In this work we
additionally consider the data transfers between L2 and L1
and between L1 and the registers, which may become the
bottlenecks instead in applications with effective cache reuse.

The four limiters considered here are  a reasonable selection out
of a wide range of options.
They yield a more differentiated performance ranking
by finding performance deviations between among variants which
the roof{}line model would consider equivalent.
The selected set of limiters is general enough to be applicable
to all common GPU architectures, the
underlying these mechanisms can be reasoned about without too much
architectural inside knowledge, and the associated metrics can be
accurately estimated using high-level methods, as shown in the next
section.

\section{Metric Estimation}

\subsection{Floating Point Execution}

The number of floating-point operations, needed for the FP
throughput limiter, is straightforward to get from the source code.
The code generator performs common optimizations such as
constant folding and common subexpression elimination to anticipate
the changes that the compiler probably would do.
For the applications and configurations covered here, the floating-point
execution rate is never the limiter and will thus not be considered
for the rest of this paper.

\subsection{L1 Cache to Register Transfers}

\begin{figure}
  \centering
  \includegraphics[width=0.45\textwidth]{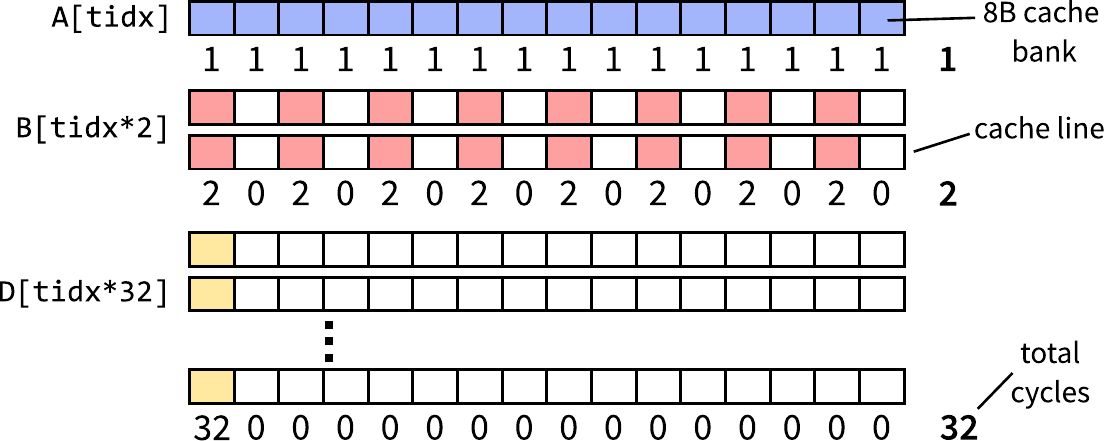}
  \caption{Illustration of cache bank conflicts}
  \label{fig:cachebanks}
\end{figure}
The number of cycles required to transfer data from the L1 cache to the registers
depends on the access patterns.
In the best case, Volta and Ampere can transfer 128B per cycle, or one unique
8B double precision (DP) value for each thread in a 16-wide half warp.
The 128B cache lines are distributed over 16
cache banks, each of which can deliver 8B per cycle. A half warp
memory access instruction can access data from different cache lines
in a single cycle, as long as at most one value per cache bank is
required.  Each additional address per cache bank requires an
additional cycle.
Consider the four loads in the following listing and the illustration
in figure \ref{fig:cachebanks}:
\begin{lstlisting}
  double a = A[threadIdx.x];
  double b = B[threadIdx.x*2];
  double d = D[threadIdx.x*32];
\end{lstlisting}
The consecutive addresses in the load from A have no bank conflicts
and result in one cycle per half warp. Load B loads only every
other 8B address, resulting in two addresses per cache bank,
which are loaded in two cycles per half warp.
The worst case is represented by the load from
D, where all addresses are in the same cache bank and have to be
loaded over 32 cycles per half warp.

We assume the total time required to transfer all data from the L1 cache to
registers to be the sum of bank conflicts of all loads.
We compute this number for each load by computing the referenced addresses
of all the threads in a half warp and counting the number of addresses per cache
bank.

\subsection{Data Volumes}

\begin{figure}
  \centering
  \includegraphics[width=0.40\textwidth]{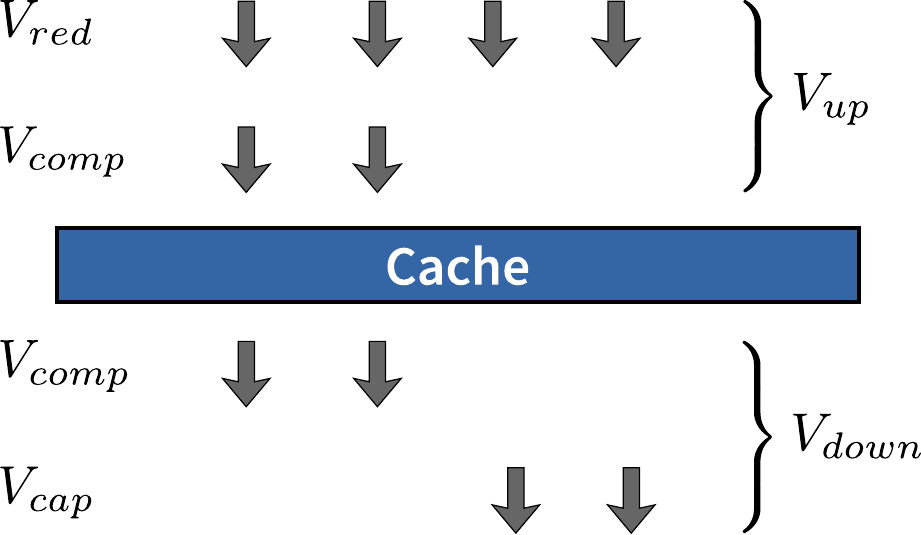}
  \caption{Illustration of in- and outgoing data volumes}
  \label{fig:capacity}
\end{figure}

A key property of memory accesses is their spatial and temporal locality.
For GPUs, temporal and spatial locality of a single thread's accesses is less relevant
than the collaborative reuse among multiple threads running on the same cache level.
It is therefore not enough to look at a single thread in isolation to determine
the amount of transferred data, but at the group of threads that share a cache level.

We denote the total volume of data transferred by a cache level (reads
and writes) due to incoming requests from the level above it as
$V_{up}$, and the volume of data caused by requests of that cache
level to the level below it as $V_{down}$ (see figure
\ref{fig:capacity}).

For memory hierarchy levels $N$ and $N-1$ (the latter being closer
to the registers), we have $V_{up}^{N} = V_{down}^{N-1}$.
$V_{down}$ can be split into three components \cite{hennessy-patterson}:

\begin{equation}
  V_{down} = V_{comp} + V_{cap} + V_{conflict}
\end{equation}

\subsection{Compulsory Misses}

$V_{comp}$ is caused by misses on the very first access to 
addresses.  Every piece of data has to be transferred into a
cache at least once for each set of threads that have the chance to
share data in a cache level.  We refer to such a set of threads as a
\emph{collaborative group} for that cache level.  The total data volume
$V_{up}$ between a cache level and the memory hierarchy level above
can be split up into the compulsory volume $V_{comp}$, which consists
of the data accessed for the first time by that collaborative group,
and the redundant volume $V_{red}$ comprising repeated requests for
data:

\begin{figure}
  \centering
  \includegraphics[width=0.5\textwidth]{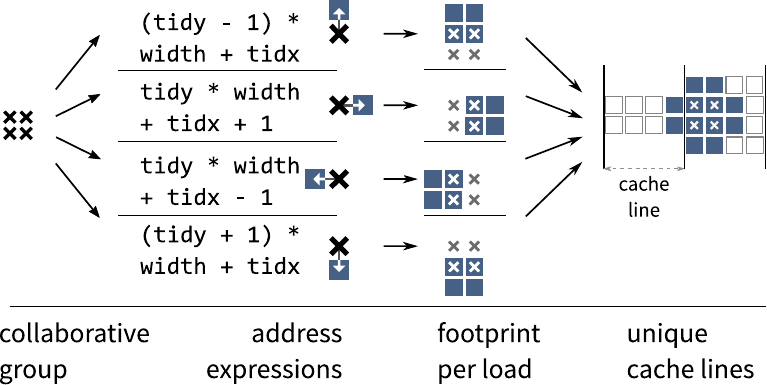}
  \caption{Illustration of the memory footprint computation. Example for a 2D 4pt stencil update and a $2\times2$ thread block.}
  \label{fig:footprint}
\end{figure}

\begin{equation}\label{eq:vup}
  V_{up} = V_{comp} + V_{red}
\end{equation}
A collaborative group's compulsory data volume $V_{comp}$ for a
cache level is its unique data footprint in that level.  The
unique data footprint is the set of unique addresses accessed by the
threads in the group at the granularity of cache lines.  To estimate
the footprint, we compute the cardinality of the set of cache lines
referenced by all threads in a representative collaborative group in
all memory accesses of a kernel. Figure \ref{fig:footprint} shows
an example where for each of the
four memory accesses, the four threads access four different
addresses.  The total amount of accessed addresses is then 16, but the
unique footprint consists of only 10 addresses since 6 are redundant.

We do this individually for each array and replace the unknown base
address of the array either by zero or by the alignment of that array.
Averaging over several representative groups makes sure that the
computed data is not an outlier caused by, e.g., being a set of
threads on a boundary by or alignment issues.
We employ two different methods to compute data footprints from addresses.

\subsubsection{Enumeration}

The first method is a direct enumeration of all addresses.  Listing
\ref{lst:griditeration} shows an example of how we use a generic grid
iteration function that can be customized with different visitors for
different applications.
The example uses a visitor that counts unique 32B cache lines to
compute the L2 load data footprint.  The different accessed fields are
counted separately, because the base addresses of the arrays are
replaced by the alignment, so that addresses of different fields could
not be kept apart.  This assumes no aliasing among different fields.
The evaluation of the thread coordinates inside a collaborative group
is vectorized using numpy's \pyth{meshgrid} function.
This explicit enumeration has the advantage of the performance being
largely independent of the complexity of the address expressions.
However, the evaluation runtime shares the $O(n\log n)$ complexity of
the \emph{unique} function regarding the number of threads in
the collaborative group.  This can become relevant for the computation
of large grids.

\begin{figure}
\begin{python}
def gridIteration(fields, threadGroup, visitor):
  for field in fields:
    for addressExpr in field.accesses:
      addresses = []
      for threadIdx in threadGroup:
        addresses.extent(addressExpr(threadIdx))
    visitor.count(addresses)

class CL32Visitor:
  CLCount = 0
  def count(addresses):
    # floor divide by 32 for 32B granularity
    # then remove duplicates
    CLCount += len(unique(addresses // 32))

def getL2LoadBlockVolume(threadBlock, fields):
  visitor = CL32Visitor()
  gridIteration(fields, threadBlock, visitor)
  return visitor.CLs * 32    #32B per cache line
\end{python}
\caption {Simplified representation of the code to compute the unique
  data footprint using a generic grid iteration visitor pattern. The
  example shows the L2 load block footprint computation. }
\label{lst:griditeration}
\end{figure}

\subsubsection{ISL}
The second method avoids this complexity by using abstract
representations of the address sets and computing these sets using the
Integer Set Library (ISL).  For example, the following definition
could be used to specify the set of all two-dimensional thread
coordinates in a $128\times 4$ thread block:
\begin{align}
\mathsf{threads} := \{[\mathsf{tidx}, \mathsf{tidy}] :\; & 0 \leq \mathsf{tidx} < 128 \; \land \\ \nonumber
                                                         & 0 \leq \mathsf{tidy} < 4\}
\end{align}
For each access to an array, a mapping from a two-dimensional thread
coordinate to a linear index is defined.  This example shows the
mapping for an access \lstinline{A[tidx, tidy+1]} to a two-dimensional
array with width $100$ and alignment $-1$:
\begin{align}
\{[\mathsf{tidx, tidy}] \rightarrow [\mathsf{idx}] : & \\ \nonumber
 \qquad \mathsf{idx} = \operatorname{floor}  ( (-1 + & \mathsf{tidx} + (\mathsf{tidy}+1)*100 ) / 4 )\}
\end{align}
The floor division converts the DP indices to 32B cache
line indices.  The ISL has functions for applying that address
mapping to the set of thread coordinates to create a set of accessed
addresses.  By creating a union of all the address sets for each
access to a particular array, a larger set with all referenced cache
lines of that array can be created.  The function
\lstinline{count_val} then computes the cardinality of that set.

The advantage of this method is the decoupling of the evaluation
runtime from the number of threads in the grid, which is good for
large thread grids.  The ISL also allows to compute more advanced set
relationships, like the intersection of two address sets, which we use
to compute the overlap of the two different data footprints.

\subsection{Capacity Misses}
\textit{Capacity Misses} happen when a redundant access misses because the data has already been evicted due to insufficient capacity hold both that piece of data and all the data accessed since the first access.
While the order of memory accesses inside a thread is fixed, the memory access order among different warps is not deterministic.
It is therefore impossible to analytically compute the exact amount of capacity misses, but it is possible to identify scenarios where capacity misses happen at all and to estimate the extent.

We define the capacity miss ratio $R_{cap}$ as the the portion of redundant accesses that will miss the cache:

\begin{equation}
  R_{cap} = \frac{V_{cap}}{V_{red}}
\end{equation}

We assume that the major factor that influences this portion is the data volume $V_{alloc}$ allocated for the collaborative group in the cache.
The allocation volume can be different from $V_{cap}$, for example in the case of Volta's L1 cache that uses a 128B for cache line allocation, but a 32B granularity for data transfers.
We define the oversubscription factor of that cache level as the ratio of the available cache capacity $V_{cache}$ and the allocation volume:
\begin{equation}
O = \frac{V_{cache}}{V_{alloc}}
\end{equation}

It is impossible to give an analytic formula $R_{cap}(O)$ for even a single application, but the behavior of that relationship in the limit can be characterized.
For an oversubscription factor less than one, there is enough cache capacity for the complete footprint and $R_{cap}$ should be zero.
At the opposite end of the spectrum for very large oversubscription, $R_{cap}$ should be large but not exceed one.

As a stand-in for a smooth transition between the two states we fit a sigmoid function of the form $\hat R(O) = a e^{-b e^{-cO}}$.
This function is not grounded in any real world mechanism, and other functions that smoothly interpolate between the two states could be used.

The capacity miss volume is then the product of the capacity miss ratio and the redundant volume, which can be computed as the difference between $V_{up}$ and $V_{comp}$:

\begin{equation}
  V_{cap} = \hat R_{cap}(O) V_{red}  = \hat R_{cap}(O) (V_{up} - V_{comp})
\end{equation}

\textit{Conflict Misses} happen when data is being evicted earlier than necessary because cache aliasing reduces the effective cache capacity.
They are note considered in this paper as a phenomenon separate from the capacity misses, as the occurrence of aliasing is to dependent on the details of an architecture's cache organization and also highly volatile.

\subsection{L2 to L1 Cache Transfers}

The relevant cooperative group for the L1 cache are the threads in a thread block, which can share data in the L1 cache of the SM they run on.
We do not consider sharing between threads of different thread blocks running on the same Streaming Multiprocessor (SM), as mostly only thread blocks that are neighbors in the thread block grid have overlap in their data footprints.
As the kernel launch execution progresses, the assignment of thread blocks to SMs becomes increasingly incoherent, and we consider the scheduling of neighbouring thread blocks to the same SM as the unlikely case.

Volta's L1 cache allocates at the granularity of 128B cache lines, but transfers data the granularity of the 32B cache line sectors.
Therefore, we compute the unique data footprint using 128B for $V_{alloc}$, but using 32B cache line sectors for $V_{comp}$.

The L1 cache uses a write-through policy, so that repeated stores to the same address have to be counted individually for $V_{down}$, so that  $V_{up} = V_{down}$ for stores.
Though loads can hit in the L1 cache on stored data, we do consider this in our estimator, as numerical code rarely mix reading and writing to the same memory locations in the same kernel.

\subsection{DRAM to L2 Cache Transfers}

The L2 cache is a chip-wide shared resource, so that all threads running on all SMs can share data in the L2 cache.
Only a comparatively small portion of the thread blocks in a kernel call grid can run concurrently.
New thread blocks are scheduled in $X-Y-Z$ order as older thread blocks complete, which results in a transient wave of running thread blocks with at sharp leading edge.
To simplify this, we subdivide the total thread block grid into discrete waves of concurrently running thread blocks.
Inside the wave, all thread blocks are considered to run simultaneously and do not have any execution order.
Considering only sharing data inside a wave, the compulsive DRAM-L2 cache volume is the unique memory footprint of all threads of that wave.

To account for the reuse of data from the previously running wave, we not only compute the unique L2 data footprint $V_{curr}$ of the current representative wave, but also compute its overlap $V_{overlap}$ with the previous wave's footprints $V_{prev}$.
This overlap $V_{overlap}$ would then represent the data that the currently running wave could hit on from the previous wave and would not have to be transferred again.
We treat capacity misses in this volume differently from other volumes, as we assume that all accesses the previous wave happen before the accesses of the current wave and will also be replaced first.
Therefore, instead of the oversubscription factor $O$, we use the coverage factor $C$ as the ratio of the previous' wave volume and the remaining L2 cache capacity after the new volume has been deducted:

\begin{equation}
C_{over} = \frac{V_{cache} - (V_{curr} - V_{overlap})}{V_{prev}}
\end{equation}

The coverage factor can become negative, if the current wave's footprint already exceeds the L2 cache capacity.

Different to the L1 cache, repeated stores are not directly written through but get cached in the L2 cache.
Because store cache lines are potentially replaced differently, we use separate fit functions $R_{cap}^{L2,store}$ and $R_{cap}^{L2,load}$.

\section{Evaluation}

\subsection{Test Environment}
All measurements are made on a NVIDIA Tesla V100-PCIe-32GB GPU using CUDA version $11.3$.
This V100 GPU consists of 80 SM's clocked at $1.38\,GHz$.
The L2 cache has a capacity of $6MB$ and the the L1 cache is configured at $128kB$.
We measured a memory bandwidth for stream kernel \textit{scale} of $790 GB/s$, and a L2 cache bandwidth of $2500 GB/s$.

\begin{figure*}[t]
  \centering
\begin{minipage}[t]{.32\textwidth}
  \includegraphics[width=\columnwidth]{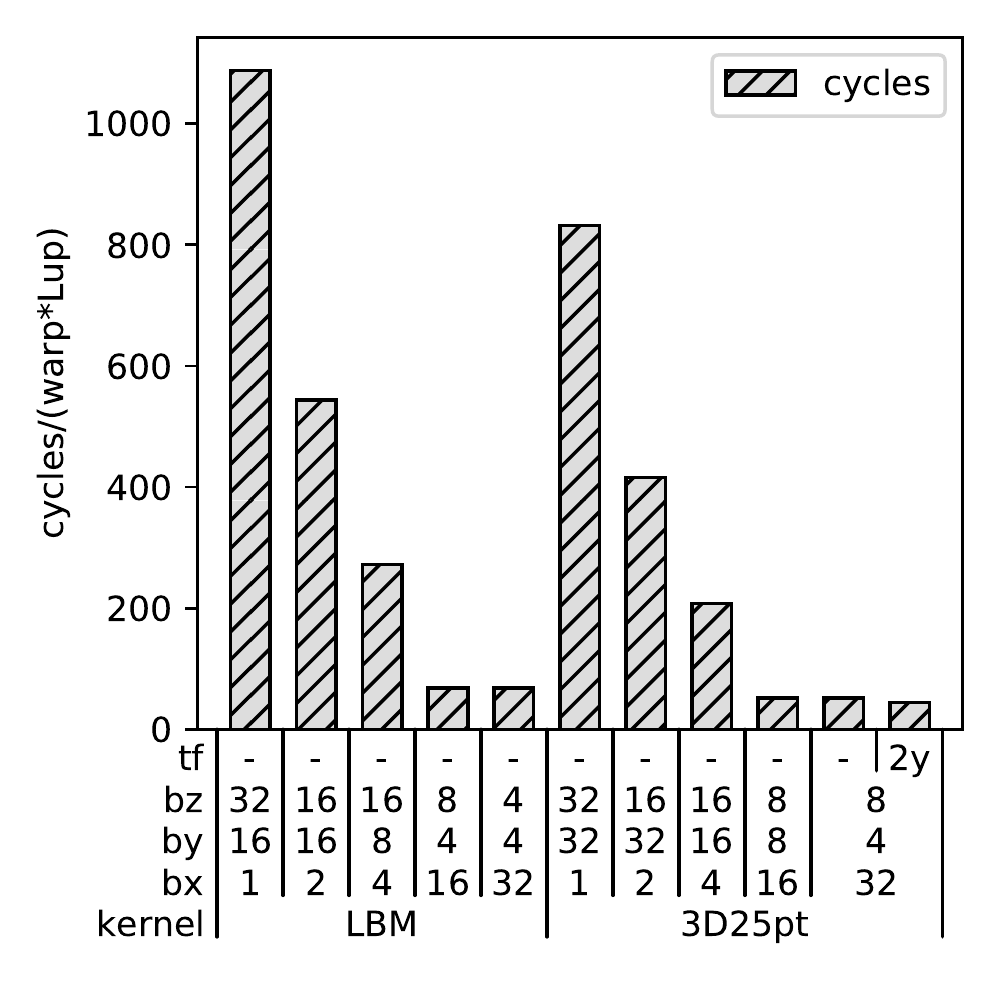}
  \caption{Estimated cycles for one lattice update of a 32 wide warp}
  \label{fig:l1thru}
\end{minipage}
\hfill
\begin{minipage}[t]{0.64\textwidth}
\begin{minipage}[t]{.49\textwidth}
  \includegraphics[width=\textwidth]{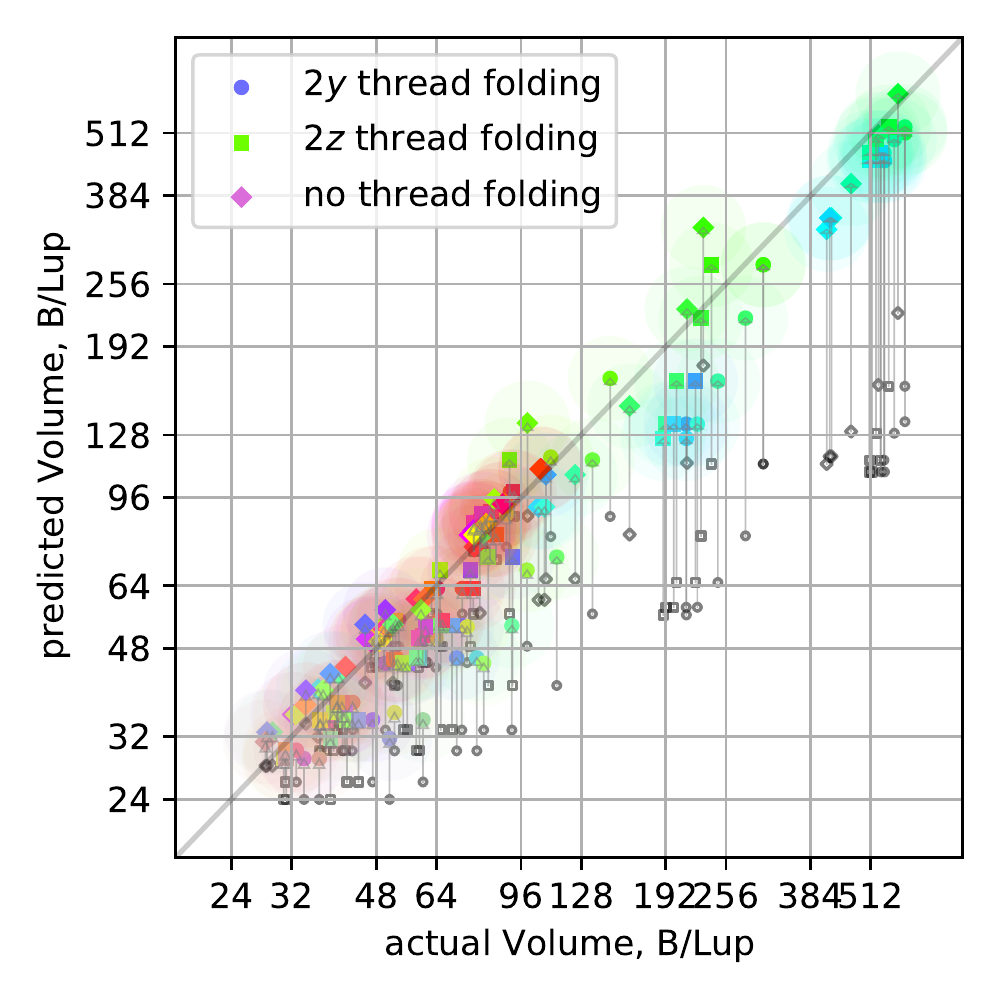}
  \caption{3D25pt/range 4 star stencil}
  \label{fig:l2loadstencil}
\end{minipage}
\hfill
\begin{minipage}[t]{.49\textwidth}
  \includegraphics[width=\textwidth]{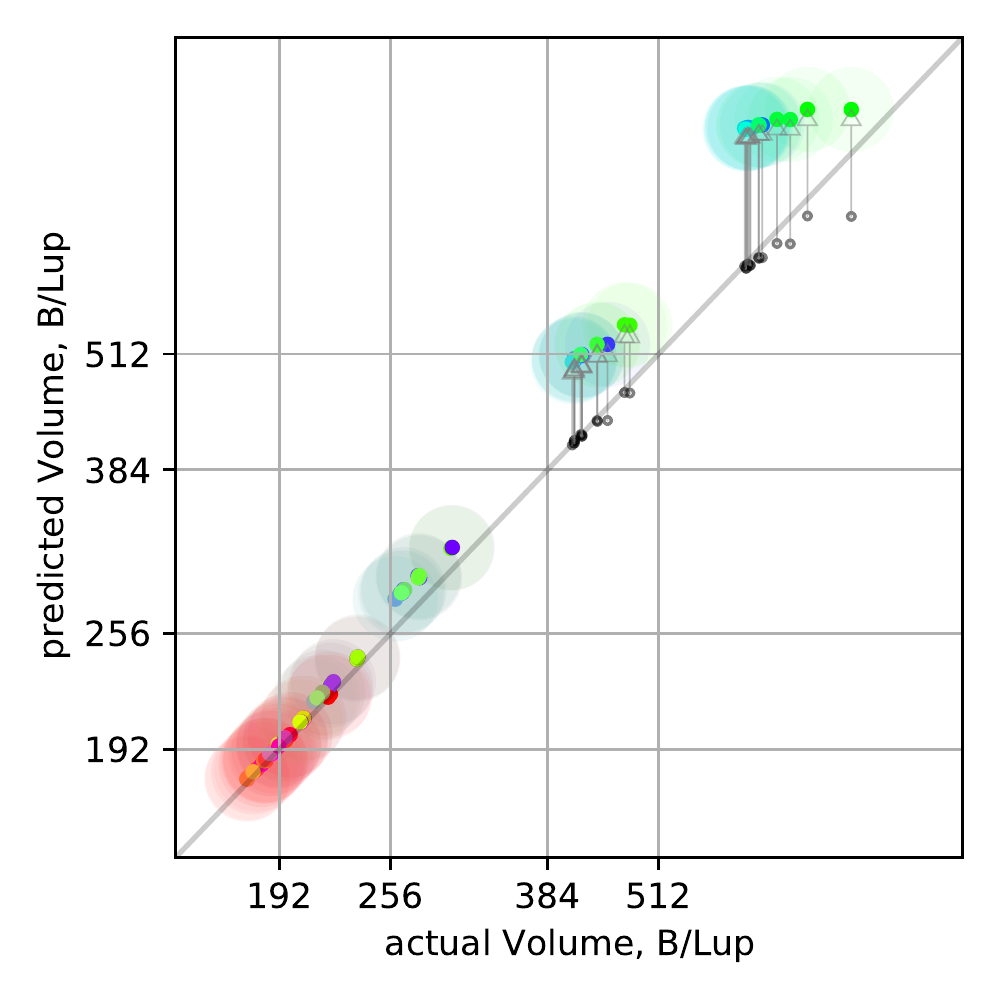}
  \caption{LBM Kernel}
  \label{fig:l2loadlbm}
\end{minipage}
  \caption*{Predicted vs measured L2-L1 loaded data volume. Data points colored by block size. Gray markers and arrows show comparison without modeling capacity misses}
\end{minipage}
\end{figure*}

\subsection{Data Points}

The accuracy of the metric estimator and the performance model is evaluated by comparing
predictions and estimates for a range of different thread block sizes  $(X,Y,Z)$ that fulfill:

\begin{align}
  & X,Y \in \{1,2,4,8,16,32,64,128,256,512\}  &\land \\ \nonumber
                 & Z \in \{1,2,4,8,16,32,64\} &\land \\ \nonumber
                 & XYZ = \begin{cases} 1024 : stencil \\ 512 : LBM \end{cases} &
\end{align}

The selection is limited so as to make sense for the application domain, and to avoid overcrowding of the graphs. The register requirements for the LBM kernel necessitates a thread block of at most 512 threads to avoid spilling.

The plots use color coded markers to differentiate different thread block sizes by mapping the three thread block coordinates to RGB colors: $RGB(\frac{log \: x}{log \:256}, \frac{log \:y }{log \:256}, \frac{log \:z }{ log \:64})$

\subsection{Long Range 3D Stencil}

The first application is a range four 3D25pt star stencil.
At 25 floating point operations and a minimum of one double precision grid point or 8B loaded and stored per lattice update (LUP), the arithmetic intensity if $~1.5 \frac{\mathsf{Flop}}{\mathsf{B}}$, is far below the machine balance of $4 \frac{\mathsf{Flop}}{\mathsf{B}}$ of a V100 for that instruction mix, which still
makes this stencil memory bound.

We use a straight forward 3D iteration scheme for the stencil, combined with
\textit{thread folding}, a technique where each thread computes multiple
consecutive grid coordinates. We test each block size with $2 \times$ thread folding
$y$ or $z$ direction as well as no thread folding.

We find that on modern GPU architectures like Volta or Ampere, other, more complex iteration schemes and optimizations for stencils, which try to deal with limitations like the comparatively small cache capacity, are not as relevant any more, as
these architectures feature much larger caches than previous GPUs.

For example, \cite{higherorderstencils} finds the straight forward 3D mapping to be the fastest on a Volta GPU, if the thread block sizes are chosen well.

The grid size for the stencil is $640\times512\times512$.

\subsection{Multi Phase LBM}

The second application is a LBM kernel using the conservative Allen-Cahn model, which can solve highly complex two-phase-flow phenomena.

We will focus on the kernel doing the interface tracking, which is one of two LB schemes that this method consists of. The curvature of the phase field is computed with a finite difference discretization, which adds a 3D7pt stencil to the conventional, very memory intensive D3Q15 LBM stencil.
This makes it particularly complicated to achieve good performance results \cite{lbmpy2}, but at the same time interesting for automatic performance estimations, which can give significant insight into the complex structure of the compute kernels.

Each lattice cell is resolved with 15 pdf values for the LB calculation, and additionally, the information of the phase-field of 6 neighboring lattice cells is needed. For propagating the pdf values to the neighboring lattice cells, a pull scheme is used. Thus the stores are aligned, and the loads are not. The D3Q15 stencil LBM part of the kernel transfers a data volume of $2*15*8B = 240B/Lup$ read and write without any reuse, compared to the just $16 - 64B/Lup$, depending on the cache effectiveness, for the finite-difference stencil.

\subsection{L1 Load Volume}

Figure \ref{fig:l1thru} shows the prediction of how many cycles each warp occupies a SM's L1 cache for a single lattice update.
For the access patterns of the two kernels, the thread block width is also the width of the contiguous blocks of accessed memory.
For 16 threads or higher, each half-warp of 16 threads loads a single contiguous block without any bank conflicts, that can be loaded in a single cycle.
With each decrease in the thread block width, the number of bank conflicts increases.
The L1 time per lattice update can be decreased through thread folding, which is shown for one data point and $2\times$ thread folding in the y-direction, because values can be reused from registers instead of reloading them.

\subsection{L2-L1 Data Volume}

\begin{figure*}[t]
  \centering
\begin{minipage}[b]{.65\textwidth}
  \includegraphics[width=\textwidth]{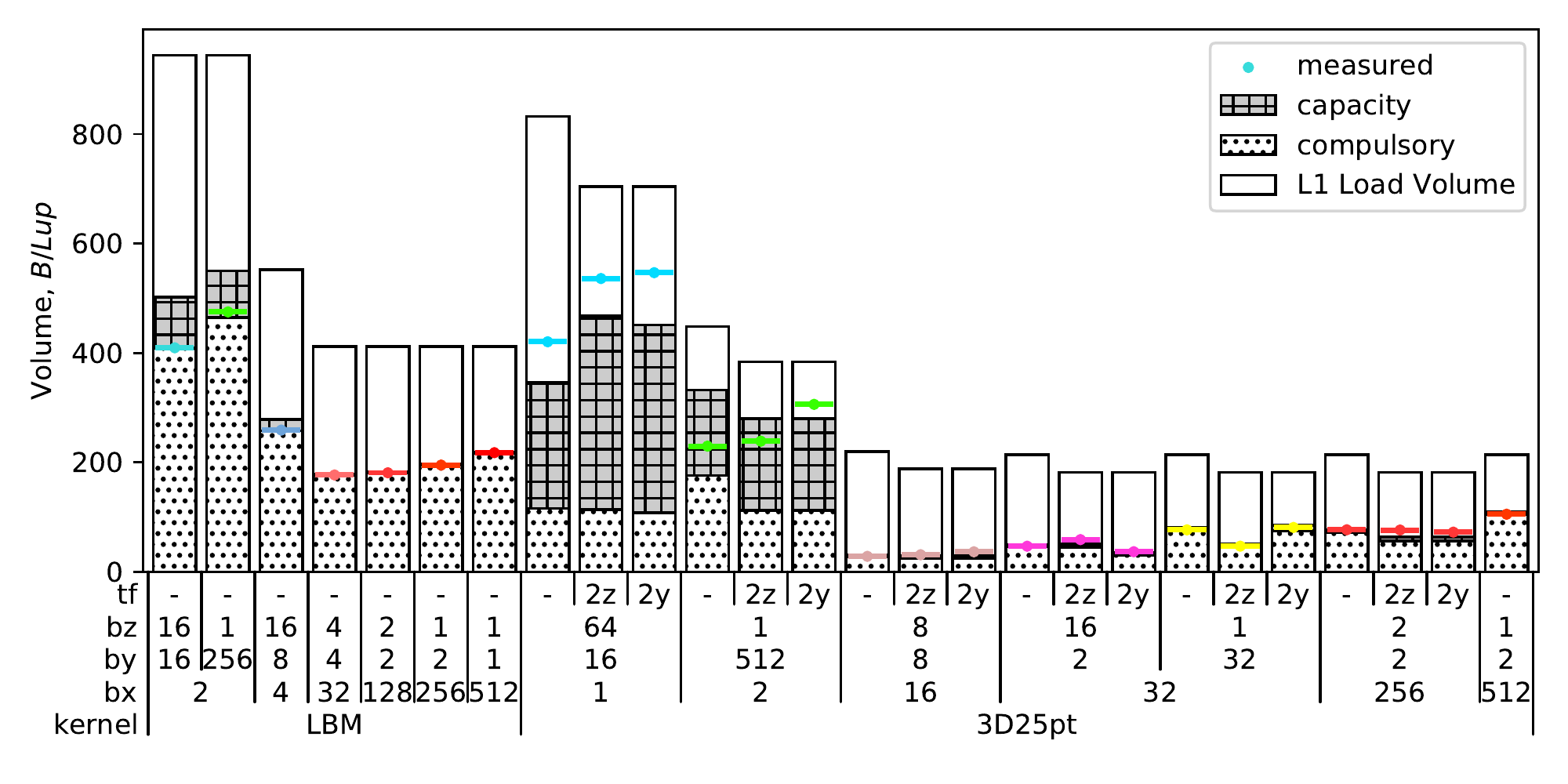}
  \caption{Composition of the L2-L1 load data volume for selected block sizes}
  \label{fig:l2comp}
\end{minipage}
\hfill
\begin{minipage}[b]{.32\textwidth}
  \includegraphics[width=\textwidth]{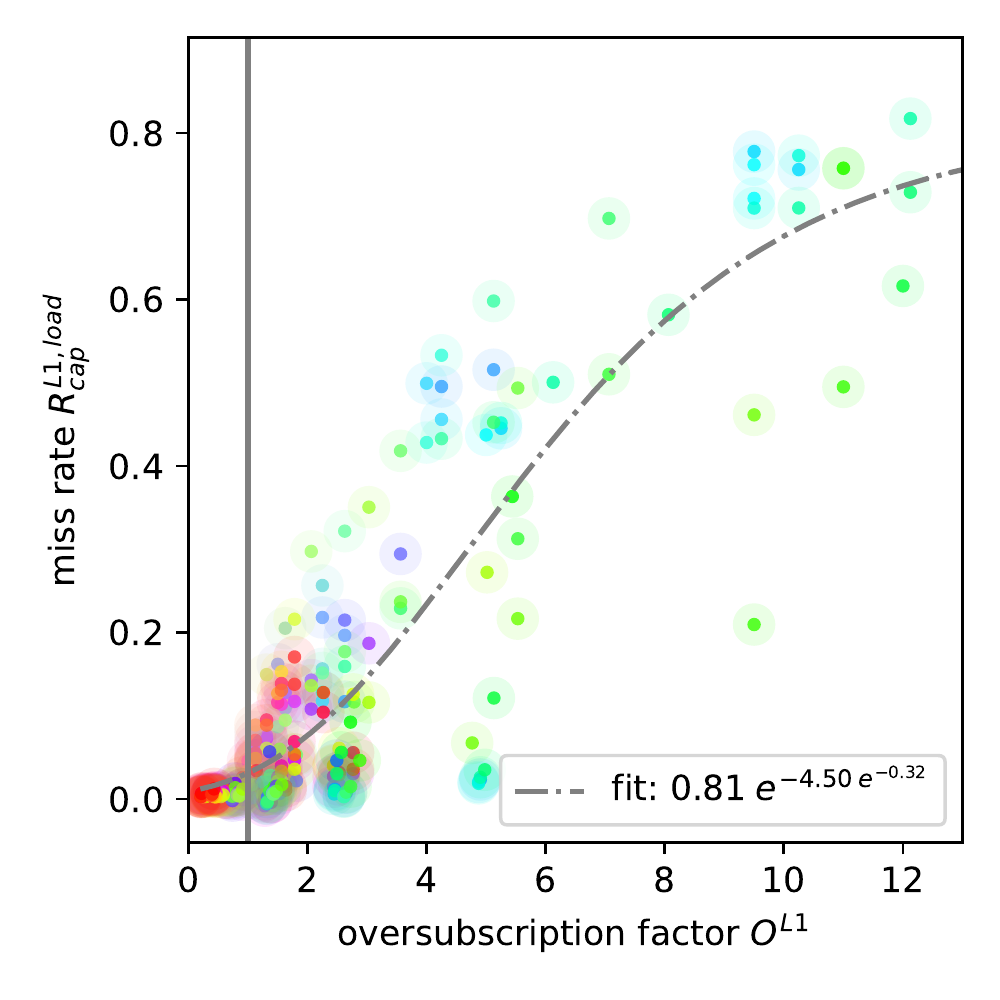}
  \caption{$R_{cap}$ vs oversubscription factor}
  \label{fig:rcap_l1}
\end{minipage}
\end{figure*}

Figures \ref{fig:l2loadstencil} and \ref{fig:l2loadlbm} plot the estimate of the L2-L1 cache load data volume in comparison to the actual data volume measured using hardware performance counters.
Plot points on the diagonal would be block sizes where the estimate is identical to the actual value, whereas points in the upper-left / lower-right triangular area of the plot are over / under estimated.

In figure \ref{fig:l2comp}, these predictions are broken down into compulsory and capacity reason for a selected number of thread block sizes.
The shown L1 cache volume is an upper limit on the L2-L1 volume, and would be completely filled by the capacity miss bar if $R_{cap} = 1$, and empty if $R_{cap} = 0$.

For the stencil, the different thread block sizes cause a wide range of L2-L1 cache data volumes.
Pale, low saturation colors, representing thread block sizes with balanced, cube shaped dimensions, e.g. $(16,8,8)$, have the lowest volumes due to their low surface volume ratio.
For the same reason, thread block sizes with one short dimension, e.g. $(512,2,1)$ or $(2, 512, 1)$, colored in more saturated primary colors, have higher data volumes.
A very short x dimension e.g. $(1, 16, 64)$, represented with colors without red, i.e. green and blue, lead to both inefficient use of the transferred 32B cache line sectors, and also inefficient allocation of 128B cache lines and consequently, large amount of capacity misses.

Thread folding in the right dimension leads to lowered compulsory volumes, but also to increased capacity misses.
The majority of the data points that have capacity misses are using thread folding.

The difference in the compulsory volumes is less pronounced for the LBM kernel, due to the high amount of streaming data volume.
The streaming component favors a x dimension as large as possible, as this minimizes the percentage of partially used cache lines used by unaligned loads.
As a compromise with the preference of the stencil part for cubed shapes, thread block sizes like $(32,2,2)$ show the lowest volumes here.
The amount of capacity misses is generally very low.
We assume that the non-LRU replacement policy of Volta's L1 cache replaces data with streaming access patterns preferrentially, so that the data of the stencil component is not displaced by the large data volume loaded by the LBM part.

As a future extension, it would be possible to automatically classify allocated data volume into streaming, i.e. a single reference, and reuse volume to differentiate the two scenarios.

Figure \ref{fig:rcap_l1} plots the measured $R_{cap}$ against the oversubscription factor of the L2 cache, and the sigmoid fit function that we use to estimate $R_{cap}$.

\subsection{DRAM-L2 Data Volume}

\begin{figure*}[t]
  \centering
\begin{minipage}[t]{.32\textwidth}
  \includegraphics[width=\textwidth]{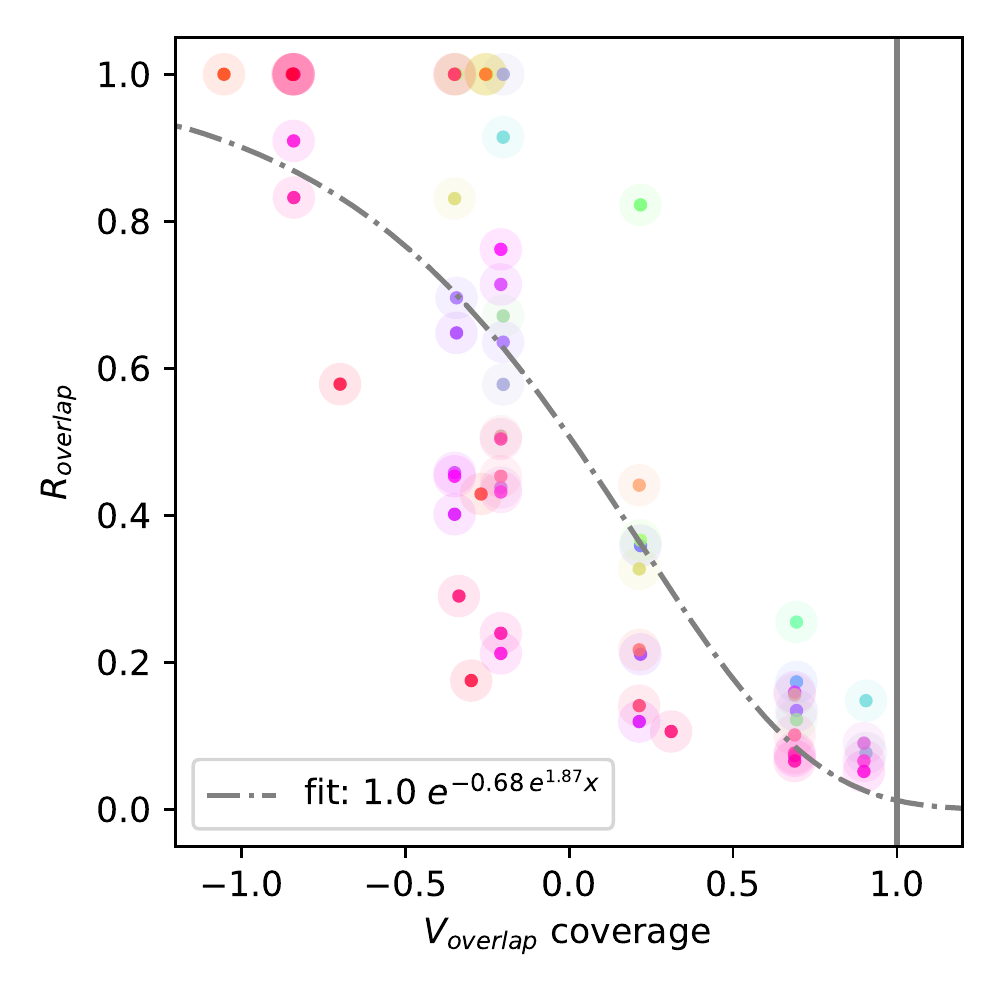}
\caption{Overlapping data volume $V_{overlap}$}
  \label{fig:rovermiss}
\end{minipage}\hfill
\begin{minipage}[t]{.32\textwidth}
  \includegraphics[width=\textwidth]{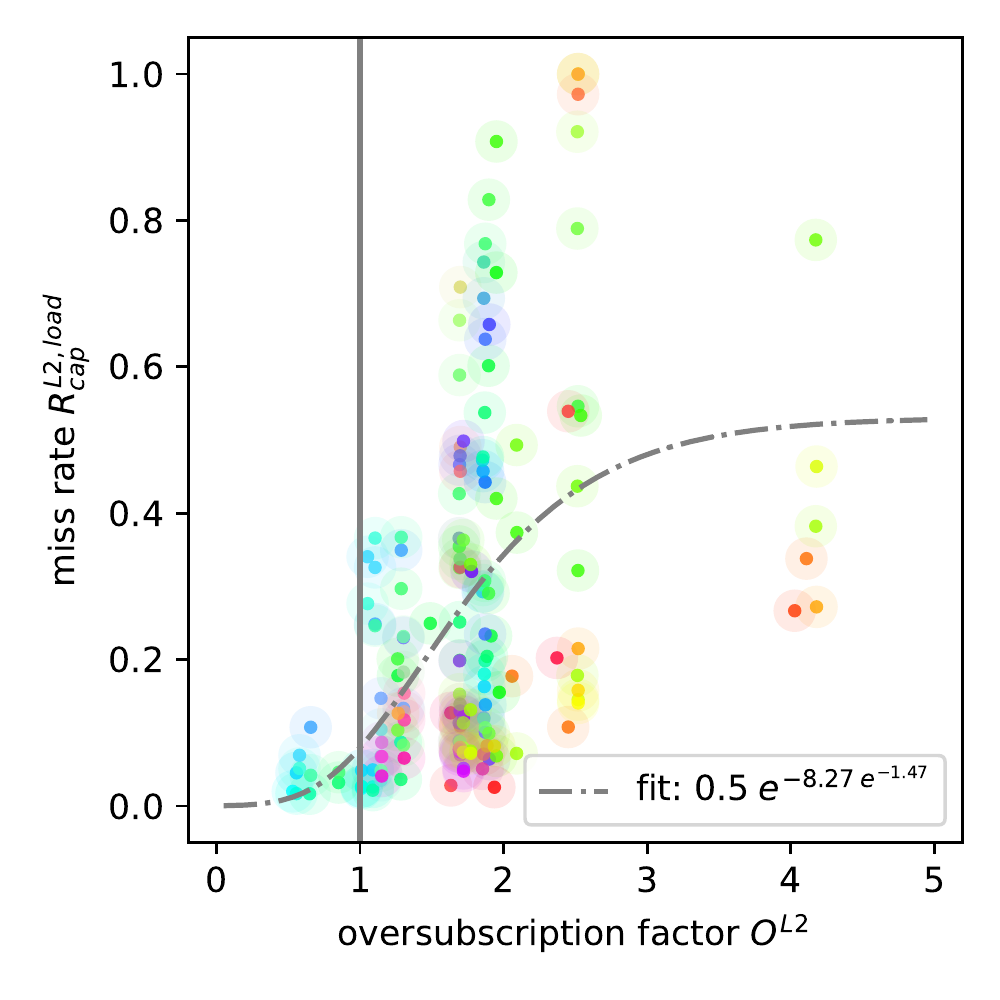}
  \caption{Redundant data volume $V_{red}$}
  \label{fig:rmiss}
\end{minipage}
\hfill
\begin{minipage}[t]{.32\textwidth}
  \includegraphics[width=\textwidth]{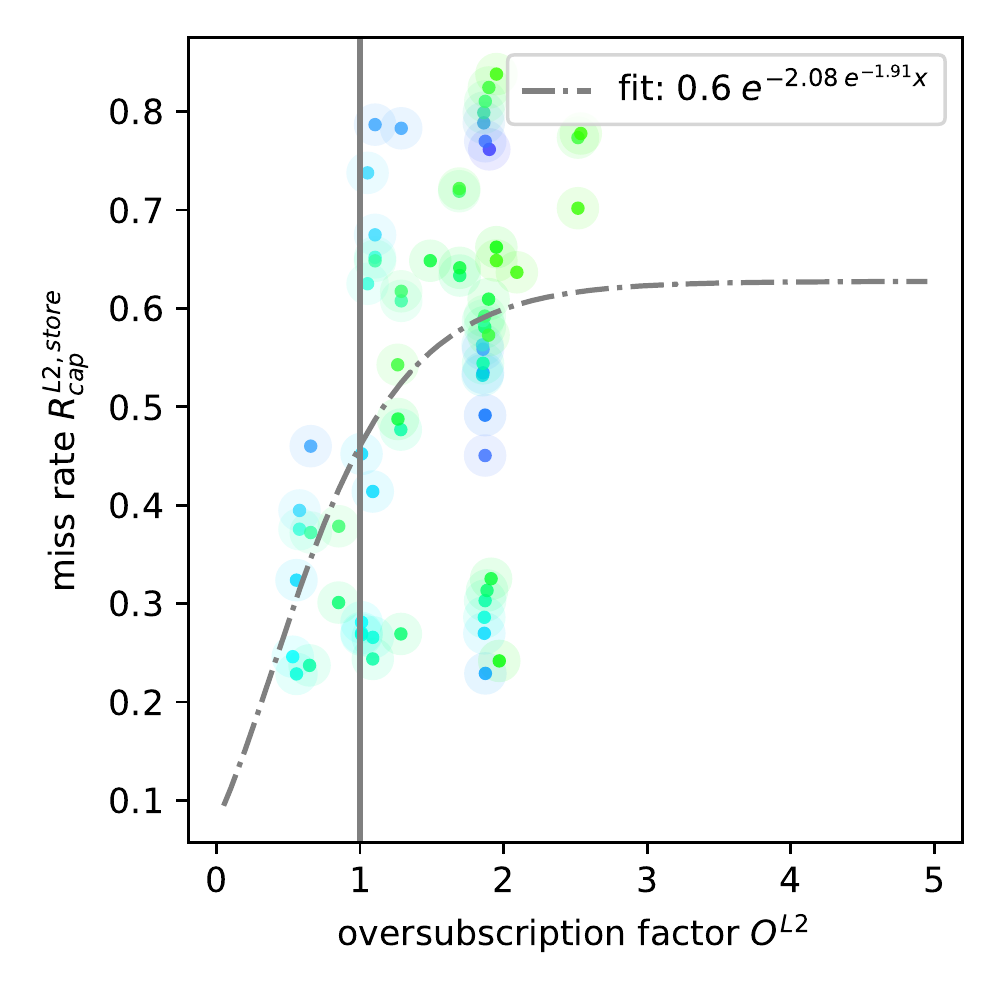}
  \caption{Repeated stores to the same cache line}
  \label{fig:rstoremiss}
\end{minipage}
\caption*{Capacity L2 cache miss rates against the oversubscription factor / the cache coverage and the used fit functions}
\end{figure*}

\begin{figure*}[t]
  \centering
\begin{minipage}[t]{.65\textwidth}
  \includegraphics[width=\textwidth]{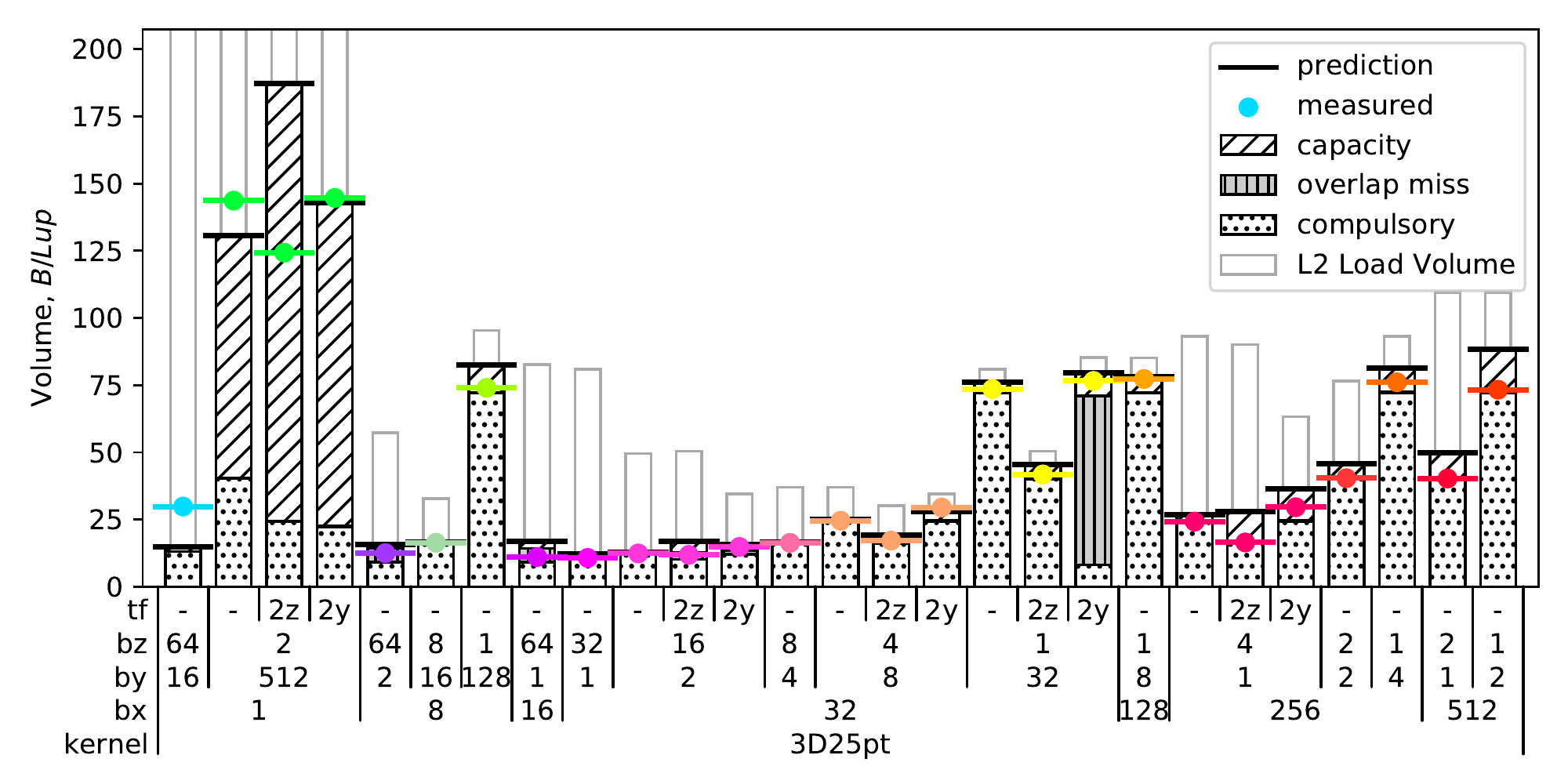}
  \caption{}
  \label{fig:compmemstencil}
\end{minipage}\hfill
\begin{minipage}[t]{.34\textwidth}
  \includegraphics[width=\textwidth]{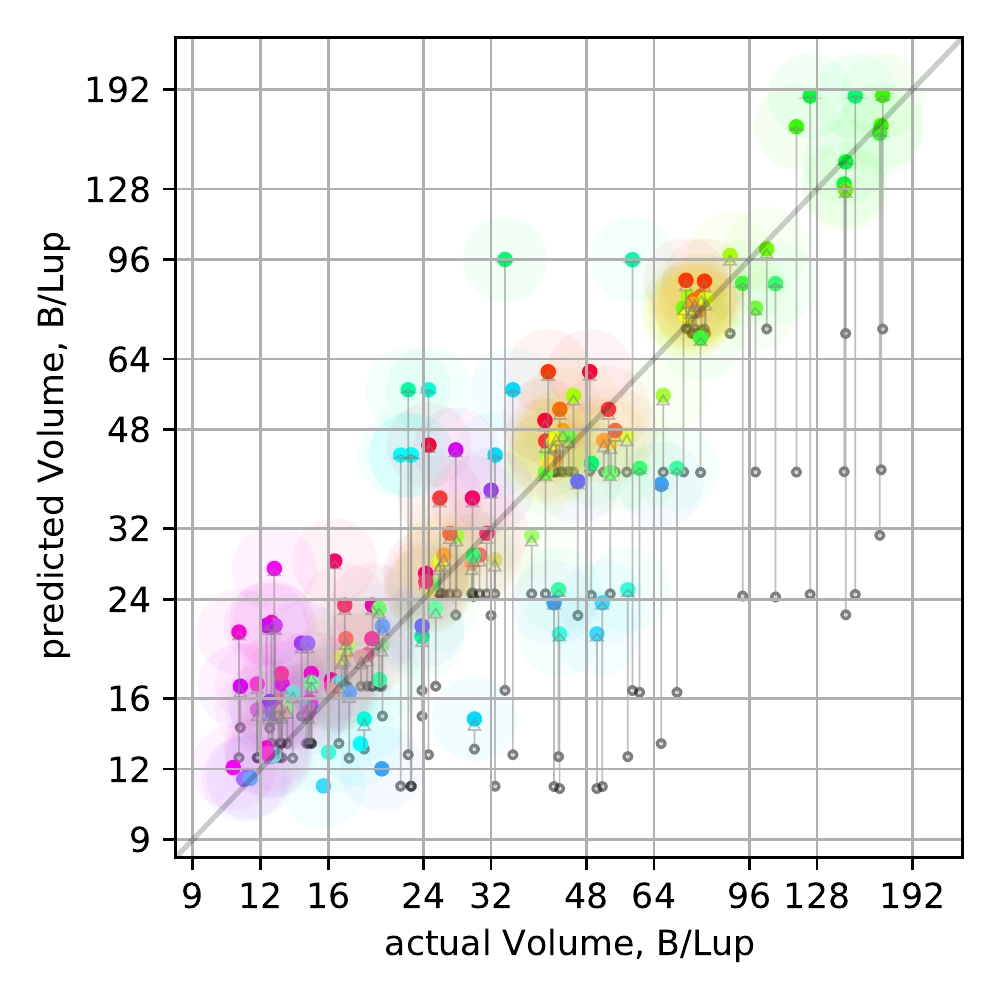}
  \caption{}
  \label{fig:memstencil}
\end{minipage}
\caption*{Left: Composition of stencil DRAM load data volumes for selected block sizes $(bx, by, bz)$ and thread folding factors ($2y$, $2z$ or no folding). Right: Prediction vs measurement of DRAM load data volumes for the long range star stencil. Gray comparison markers show the change when including capacity misses.}
\end{figure*}

\begin{figure*}[t]
  \centering
\begin{minipage}[t]{.65\textwidth}
  \includegraphics[width=\textwidth]{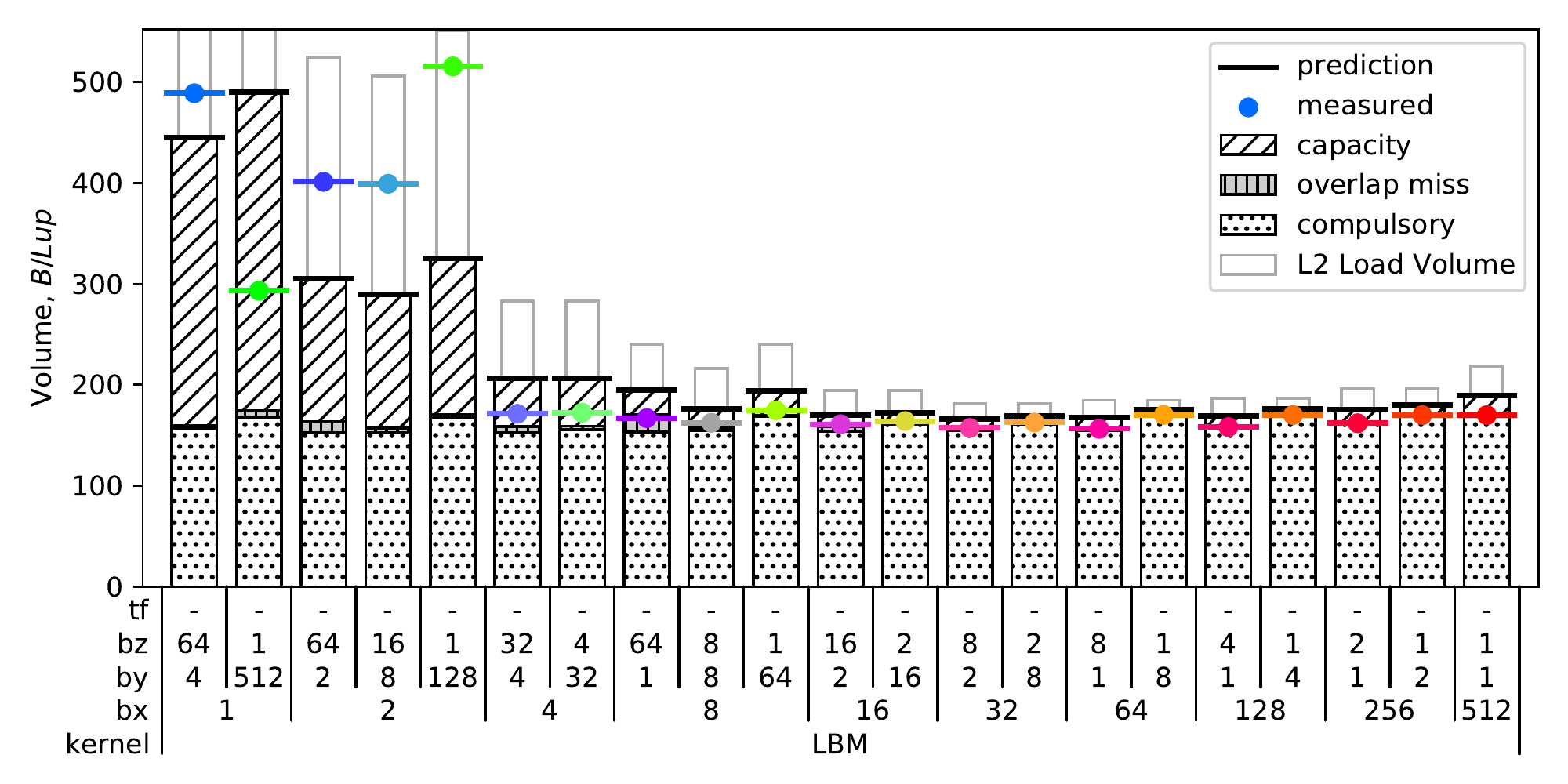}
  \caption{}
  \label{fig:compmemlbm}
\end{minipage}\hfill
\begin{minipage}[t]{.32\textwidth}
  \includegraphics[width=\textwidth]{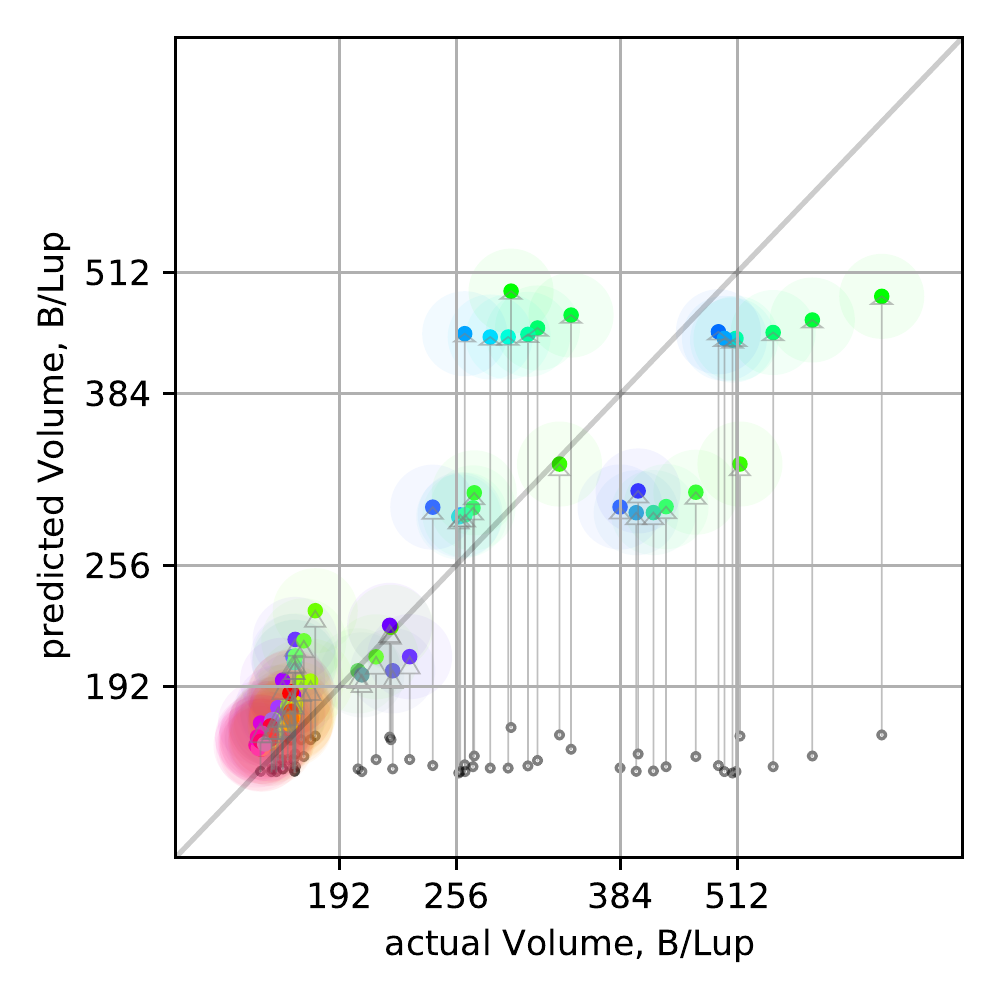}
  \caption{ }
  \label{fig:memlbm}
\end{minipage}
\caption*{Left: Composition of stencil DRAM load data volumes for selected block sizes $(bx, by, bz)$. Right: Prediction vs measurement of DRAM load data volumes for the LBM kernel. Gray comparison markers show the change when including capacity misses.}
\end{figure*}

The factor $R_{overmiss}$ can be derived from a memory volume measurement by deducting the minimal compulsive data volume excluding the overlapping data volume from the measured data volume and dividing the difference by overlapping volume.
This derived overlapping data miss ratios is plotted in Figure \ref{fig:rovermiss} against the the previous wave coverage, together with the fit of a sigmoid function that we use to model $R_{overmiss}$.
Although there is a wide variability in the miss rate for the different data points and deviations from the fit function, it still captures the clear transition from a low miss rate for full coverage to high miss rates for an oversubscribed L2 cache.

The values for $R_{cap}^{L2,load}$ and $R_{cap}^{L2,store}$ can be similarly derived from the measurements by deducting the compulsory volumes and overlap capacity misses from the measured data volumes, the results of which are plotted in Figures \ref{fig:rmiss} and \ref{fig:rstoremiss} against the oversubscription factor.
Both ratios again are not a simple function of the oversubscription factor, but show a transition from low capacity miss rates at sufficient coverage to a range of miss rates at higher oversubscription, which we represent with the shown sigmoid fit functions.

Figure \ref{fig:memstencil} shows a comparison of the memory load volume predictions of all stencil data points against the measured volumes.
The plot shows that thread blocks can be sorted by color into three categories, which are also represented in the detailed breakdown by component in figure \ref{fig:compmemstencil}.
Purple or violet colors, i.e. wide and deep thread block sizes with large $x$ and $z$ dimensions, e.g. $(32,1,32)$, have the lowest volumes.
Yellow/orange colors, representing wide and tall thread block sizes with large $x$ and $y$ dimensions, e.g. $(32,32,1)$, have larger data volumes.
Green, tall thread blocks with a large $y$ component, e.g. $(2,512,1)$, have the highest data volume.

The deeper thread block sizes, i.e. large z dimension, result in a lower balance because
the grid is filled with thread blocks in $x,y,z$ order.
Most of the time a wave consists only of one layer of thread blocks in the $z$ direction, making the $z$ extent of the whole wave entirely dependent on the depth of a single thread block.
A very shallow wave, i.e. low $z$ extent,  results in little reuse in $z$ direction and hence high volumes.

The capacity miss effects are largest for tall blocks with a large $y$ component and short $x$ component.
These block sizes combine large L2 cache data allocations with many redundant L2-L1 data transfers, each of which is a potential capacity cache miss.

Thread folding in the right dimension can reduce the compulsory data volume, see e.g. $(32,32,1)$.
A $2y$ thread folding is counterproductive here, as the $y$ component of the wave was already large.
The resulting large overlap with the previous wave does not come to fruition due to the increased L2 cache allocation and correspondingly increased capacity miss rates.

The $2z$ thread folding instead doubles the $z$ extent of the wave from one to two, which improves the surface volume ration of the wave and leads to a smaller compulsory volume.

Figures \ref{fig:memlbm} and \ref{fig:compmemlbm} show the same data for the LBM kernel.
The streaming nature of the LBM part of the kernel leads to a much smaller dependence on the $z$ extent than the stencil.
Instead, the memory data volumes for the LBM kernel are mostly influenced by the thread block x dimension.
The shorter the $x$ extent of a thread block, the fewer complete cache lines are loaded by a thread block in comparison to the partial cache lines loaded at the thread block boundary due to the unaligned LBM component loads, and the larger the amount of redundant loads.
This is even more valid for $x=1,2$, where all loads access only partial cache lines.
This is reflected in the component breakdown in figure \ref{fig:compmemlbm} by decreasing L2 load volumes with increasing $x$ extent.
Combined with increased L2 cache allocation when a wave does not fill a $x$ row completely, this leads to capacity misses for small $x$ extents.

\subsection{Performance}

\begin{figure*}[t]
  \centering
  \hfill
  \begin{minipage}[t]{.39\textwidth}
  \includegraphics[width=\textwidth]{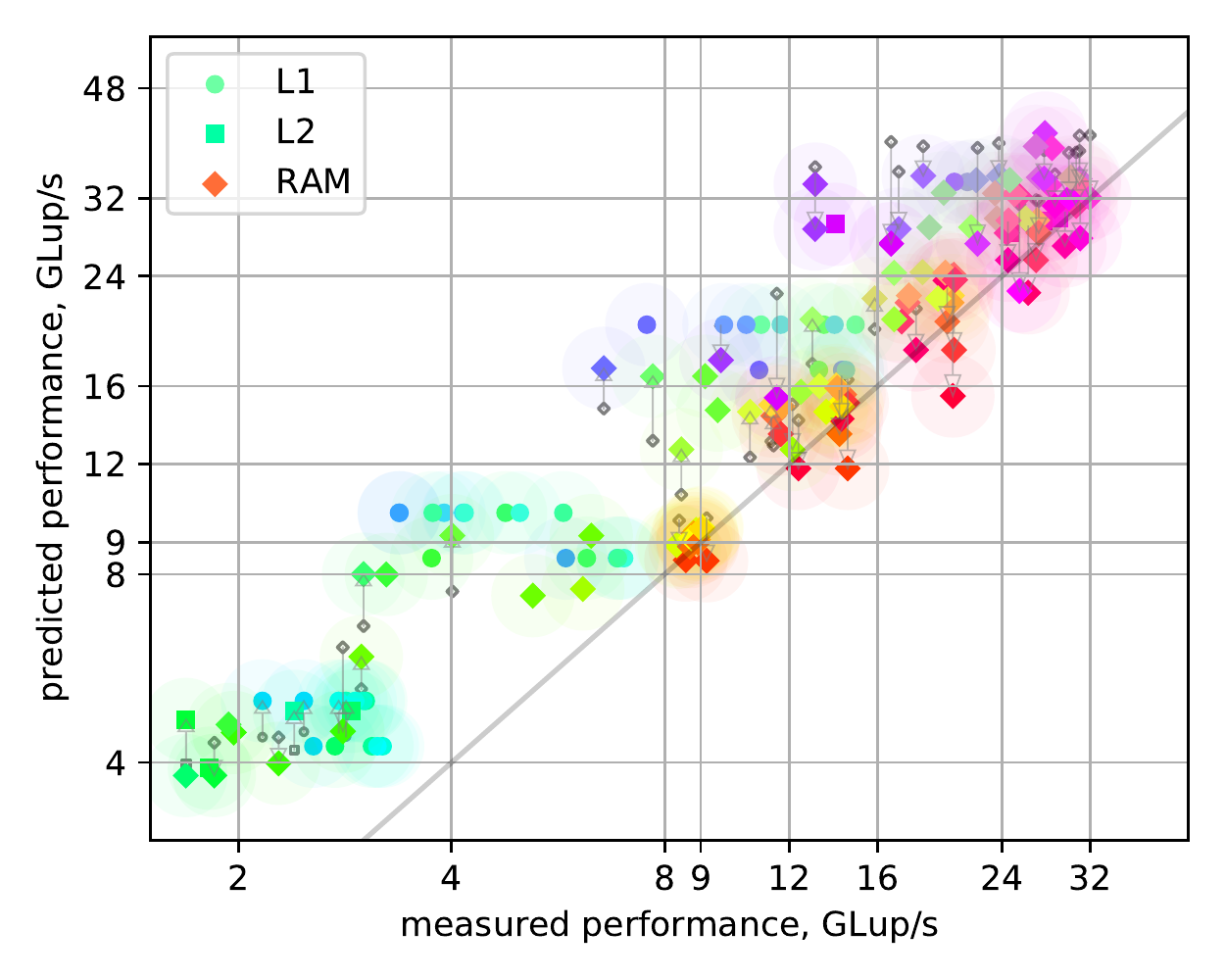}
\caption{3D25pt/range 4 star stencil}
  \label{fig:rooflinestencil}
\end{minipage}\hfill
\begin{minipage}[t]{.39\textwidth}
  \includegraphics[width=\textwidth]{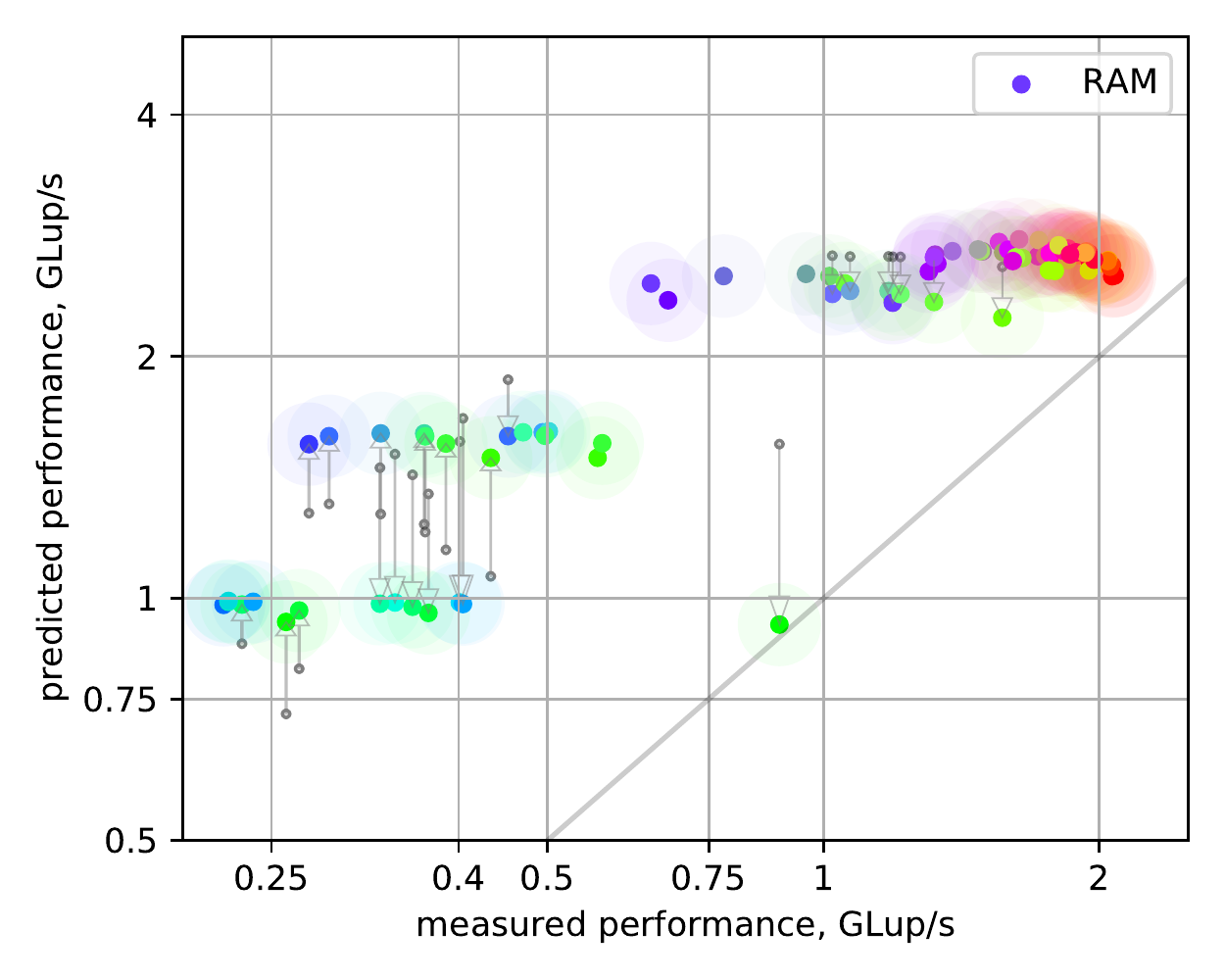}
  \caption{LBM kernel}
  \label{fig:rooflinelbm}
\end{minipage}
\hfill
\begin{minipage}[b]{0.13\textwidth}
  \caption*{Performance prediction vs.\ measurement. The gray comparison
    markers and arrows show the difference to a phenomenological
    prediction made using measured data.}
\end{minipage}
\end{figure*}

Figures \ref{fig:rooflinestencil} and \ref{fig:rooflinelbm} show
comparisons of the predicted performance using the estimated data
volumes as input for the presented performance model.  The gray
comparison markers show the difference to a phenomonological
prediction that uses the same performance model but measured
data volumes.
For both applications, the performance model shows
overprediction, regardless of whether estimated or measured
data volumes are used.

For the 3D25pt range-four star stencil, the predictions are able to
rank the different configurations by performance and clearly captures
the performance differences between well-performing and badly-performing
configurations.  The configuration with the best predicted
performance, $(16, 2, 32)$ without thread folding, is the 13th best
out of 162 configurations at $86\%$ ($27.6 GLup/s$) of the
performance compared with the fastest measured configuration,
$(32, 2, 16)$ with $2z$ thread folding at $31.9 GLup/s$.  The
inability to find the actual fastest configuration is not due to
inaccurate data volumes, as the phenomenological performance model
using measured data volumes picks the same configuration as the
fastest.  The performance model cannot resolve 
the differences among the collection of well-performing configurations.
It does, however, correctly identify the
general type of configuration that performs well, as the best-predicted
and best-measured configuration do have similar shapes.

This is a relevant accomplishment, which is illustrated by comparison with
the thread block sizes found in \cite{higherorderstencils}. They find that among the
sizes $(8,8,8)$, $(16,16,4)$ and $(32,32,1)$, the
first one performs best for the application of a very similar
stencil to the one we used.  However, we find that in our
measurements, the best predicted thread block size $(16,2,32)$ is
$36\%$ faster than a $(8,8,8)$ thread block size.  It is a nonintuitive
insight that a $(16,2,32)$ thread block size performs better
than the $(8,8,8)$, $(16,16,4)$ or $(32,32,1)$ thread block sizes they
have selected.

For this stencil, the most important limiter, especially with the
fastest configurations, is the DRAM bandwidth.  Although fewer
configurations are limited by the L2 cache bandwidth, the thread block
sizes with the lowest DRAM balance like $(32,1,32)$ are L2 cache
bandwidth limited because of their flat shape.  The L1 cache
limitation comes into play only for thread block sizes with very small
$x$ dimensions.

For the LBM kernel, the performance model manages to correctly
identify the worst-performing configurations with short $x$
dimensions.  Apart from that, it cannot distinguish the averagely
performing from the well performing configurations.  Just as with the
stencil kernel, the fault is not with the estimated data volumes, but
with the performance model, which does not capture the relevant
mechanisms here.  The LBM kernel, due to its streaming nature, is
limited entirely by the DRAM bandwidth.

\section{Conclusion and Outlook}

We have demonstrated an automated performance modeling process
for loop kernels on GPUs that is based on tracking data
accesses via address expressions. 
Our method can
estimate the data volumes transferred between the levels
of the memory hierarchy with high accuracy.  Its versatility
for a wide range of GPU programs has been demonstrated by
evaluating it with two challenging and diverse kernel types, a
long-range star stencil and a complex LBM kernel with a mix of access
characteristics.

We have shown the usefulness of these data volumes to gain insight
into the performance characteristics of a program and to classify
different code generation configuration by their performance.
However, our evaluation also showed that the simple performance model
does not capture all the performance relevant mechanics and fails to
differentiate between configurations at the top of the ranking.
Identifying and modeling these mechanisms is an important topic for
future work.
Modeling Translation Lookaside Buffer (TLB) misses would be one of the
candidates, where the relevant program metric would be TLB pages
accessed by the current wave.

Another topic of future work is the testing and extension for
different hardware architectures.  The general hardware model with a
local L1 cache and a shared L2 cache is applicable for all current GPU
architectures, but details like cache line sizes and cache
capacities have to be adapted.  For example, AMD's current CDNA
architecture's much smaller L1 cache would lead to many more capacity
misses.  On NVIDIA's Ampere, the much larger L2 cache is
split, which leads to traffic between the two halves.

We are also looking to verify the applicability of our
method for more applications and more complex code transformations
like temporal blocking.
For temporal blocking and other complex stencil iteration schemes,
our solution could be used to choose parameters
like blocking factors and parallelization schemes.

\bibliographystyle{IEEEtran}
\bibliography{references}

\end{document}